\documentclass[runningheads,a4paper]{llncs}

\usepackage{amssymb}
\usepackage{graphicx}
\usepackage[caption=false]{subfig}
\usepackage{booktabs}
\usepackage{multirow}
\usepackage{adjustbox}

\usepackage{url}
\urldef{\mailsa}\path|{alfred.hofmann, ursula.barth, ingrid.haas, frank.holzwarth,|
\urldef{\mailsb}\path|anna.kramer, leonie.kunz, christine.reiss, nicole.sator,|
\urldef{\mailsc}\path|erika.siebert-cole, peter.strasser, lncs}@springer.com|

\usepackage{xcolor}

\begin{document}

\mainmatter  

\title{3D Probabilistic Segmentation and Volumetry from 2D projection images}
\titlerunning{3D Probabilistic Segmentation and Volumetry from 2D projection images}

%
%
\author{Athanasios Vlontzos
\and Samuel Budd
\and Benjamin Hou
\and Daniel Rueckert
\and Bernhard Kainz}

\authorrunning{A. Vlontzos et al.}

\institute{BioMedIA\\ Imperial College London\\ London, UK
}

\toctitle{Lecture Notes in Computer Science}
\tocauthor{Authors' Instructions}
\maketitle

\begin{abstract}
X-Ray imaging is quick, cheap and useful for front-line care assessment and intra-operative real-time imaging (e.g., C-Arm Fluoroscopy). However, it suffers from projective information loss and lacks vital volumetric information on which many essential diagnostic biomarkers are based on.
In this paper we explore probabilistic methods to reconstruct 3D volumetric images from 2D imaging modalities and measure the models' performance and confidence. 
We show our models' performance on large connected structures and we test for limitations regarding fine structures and image domain sensitivity. 
We utilize fast end-to-end training of a 2D-3D convolutional networks, evaluate our method on 117 CT scans segmenting 3D structures from digitally reconstructed radiographs (DRRs) with a Dice score of $0.91 \pm 0.0013$. Source code will be made available by the time of the conference. 

\end{abstract}

\section{Introduction}
Computed tomography (CT) scans provide detailed 3D information of patient anatomy that is vital to many clinical workflows. For many pathologies, accurate diagnosis relies heavily on information extracted from CT images and volumes~\cite{doi:10.1148/radiol.14141356}, e.g. biomarkers derived from 3D lung segmentations are used to characterize and predict Tuberculosis progression~\cite{VanDyck2003}. CT scans, however, are both time-consuming and expensive to perform, and are not always available at the patients current location, resulting is delayed diagnosis and treatment. CT scans also present a higher risk to the patient due to increased radiation exposure over a typical Chest X-Ray (CXR). Meanwhile CXRs are routinely taken in the clinical practice at significantly decreased cost and radiation dosage while acquisition times are many orders of magnitude less than a CT scan.

Learning based methods have shown great potential for synthesizing structurally coherent information in applications where information is lost due to non-invertible image acquisition~\cite{DBLP:journals/cgf/HenzlerRRR18}. A primary example of such an application is CXR projection. As the human anatomy is locally well-constrained, a canonical representation can be adopted to learn the anatomical features and extrapolate a corresponding 3D volume from a 2D projection view. This can be achieved by reflecting likely configurations, as they were observed in the training data, while inference is conducted by giving a sparse conditioning data sample, like a single projection. 



\noindent\textbf{Contribution:} 
We show how probabilistic segmentation techniques~\cite{Baumgartner2019,budd2019confident} can be extended with the ability to reconstruct 3D structure from projected 2D images.
Our approach evaluates the potential of deep networks to  invert projections, an unsolved problem of projective geometry and medical image analysis. 
We evaluate our method by reconstructing 3D lung segmentation masks and porcine rib-cages from 2D DRRs. We show that our approach works well for large, connected regions and test for limitations regarding fine, unconnected anatomical structures projected on varying anatomy and domain sensitivity across datasets. We further show how to adapt our methods to perform Unsupervised Domain Adaptation on NIH chest X-Rays. The proposed network is fast to train, converges within a few hours and predicts 3D shapes in real-time.

\section{Related Work}

Extracting 3D models from a single or multiple 2D views is a well-established topic in computer vision~\cite{lin2018learning,sun2018pix3d}. Earlier approaches included learning shape priors, and fitting the 3D shape model onto the 2D image. In \cite{doi:10.1080/21681163.2014.913990,koehler2009knowledge}, the authors attempt to reconstruct ribs by using, a priori known, statistical shape models.
Both methods use a bi-planar approach as they utilize 2 orthogonal X-ray views. These methods do not generate a CT like image, as they only deform a solid rib-like template. 

With the advances in deep learning, generative deep convolutional neural networks have been proposed to perform image generation in the context of medical imaging. In a recent work, parallel to ours, Ying et al. proposed X2CT-GAN~\cite{Ying_2019_CVPR} to synthesize full 3D CT volumes from 2D X-rays. Like \cite{doi:10.1080/21681163.2014.913990,koehler2009knowledge}, Ying et al. also use multiple views to create the 3D volume. However, instead of statistical shape models, Ying et al. uses generative adversarial networks (GANs) to synthesize 3D CT volumes from 2D X-rays. As GANs are trained to approximate the probabilistic distribution of the training dataset implicitly, they are known to hallucinate ``plausible'' features. This is detrimental in cases of fine structures, e.g., bronchi, blood vessels and small lesions. In the case of vessel-like structures which are almost random in construction, a GAN will hallucinate a plausible structure that is highly probable from images in the training dataset, instead of generating a structure that's extrapolated from the input. Hence the resulting structures are of poor quality, often disconnected and non realistic.

In \cite{yan2016perspective} the authors reconstruct a 3D volume of an object from a 2D image of it. Contrary to X-Rays which can be thought of as the ``shadow'' of the object, \cite{yan2016perspective} used as inputs 2D images of 3D structures, not their projections. Hence there was significantly less information loss than in the case of projections. \cite{10.1007/978-3-319-66179-7_51}~attempts a similar task to ours but aims at decomposing the provided X-Ray image rather than reconstructing the CT volume.
More aligned to our work, Henzler et al~\cite{DBLP:journals/cgf/HenzlerRRR18} creates 3D Volumes from 2D Cranial X-Rays. Their architecture  is similar to ours, however, they only regress the 3D cranial structure, whereas we attempt to regress directly to CT Hounsfield units (HU). Furthermore we adopt a probabilistic technique while their model is fully deterministic.

\section{Methodology}
\begin{figure*}[htb]
    \centering
    \subfloat[2D to 3D U-Net, Blue Blocks indicate 3D Convolutions; Orange Blocks indicate Dropout Layers, \emph{c.f.} Sect.~\ref{sect:MC}\label{fig:architecture}]{
        \includegraphics[height=4cm]{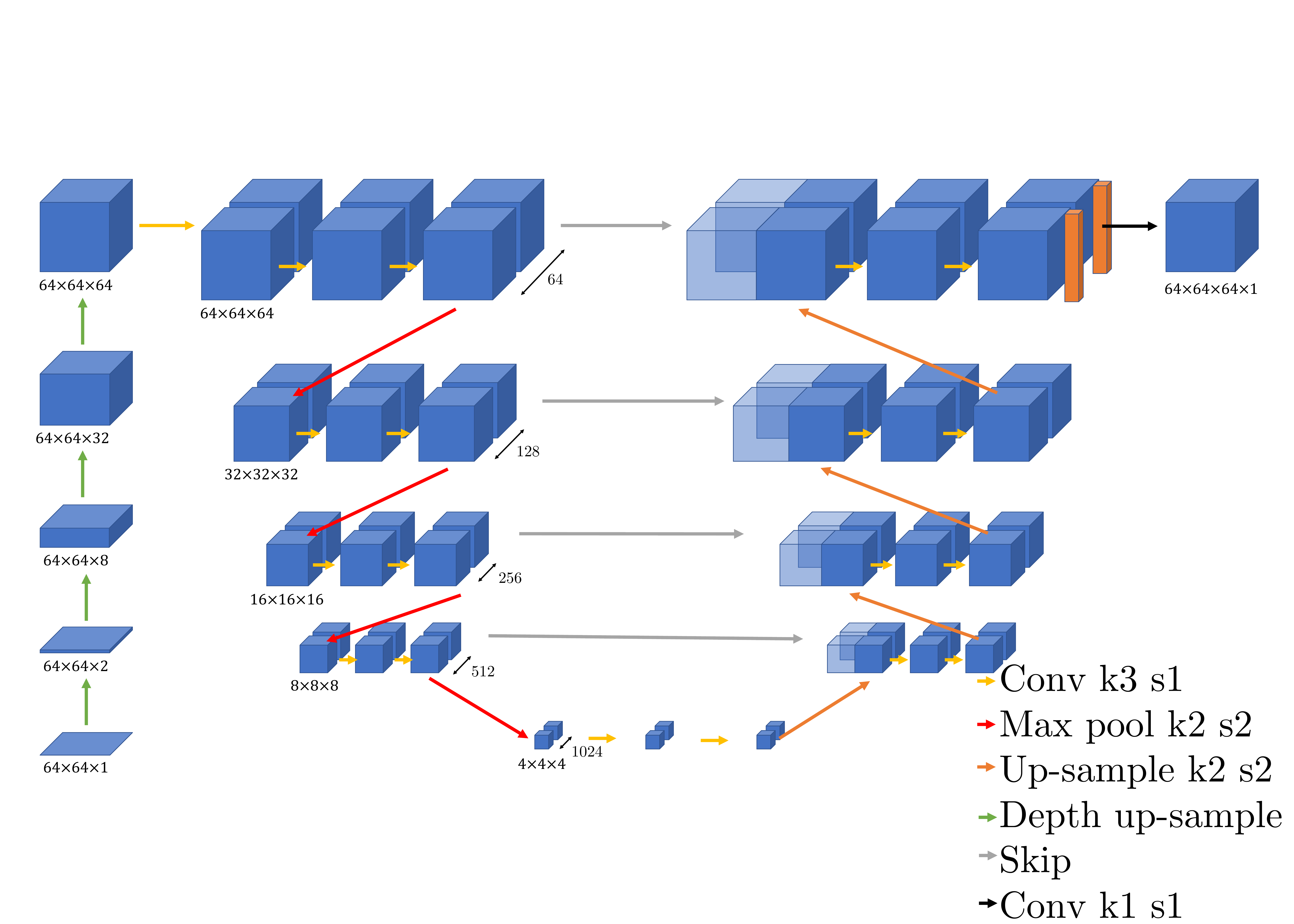}
        }
    \hfill
    \subfloat[PhiSeg\cite{Baumgartner2019} with proposed augmentations,  \emph{c.f.} Sect.~\ref{sect:phiseg}\label{phiseg_}]{
        \centering
        \includegraphics[height=4cm]{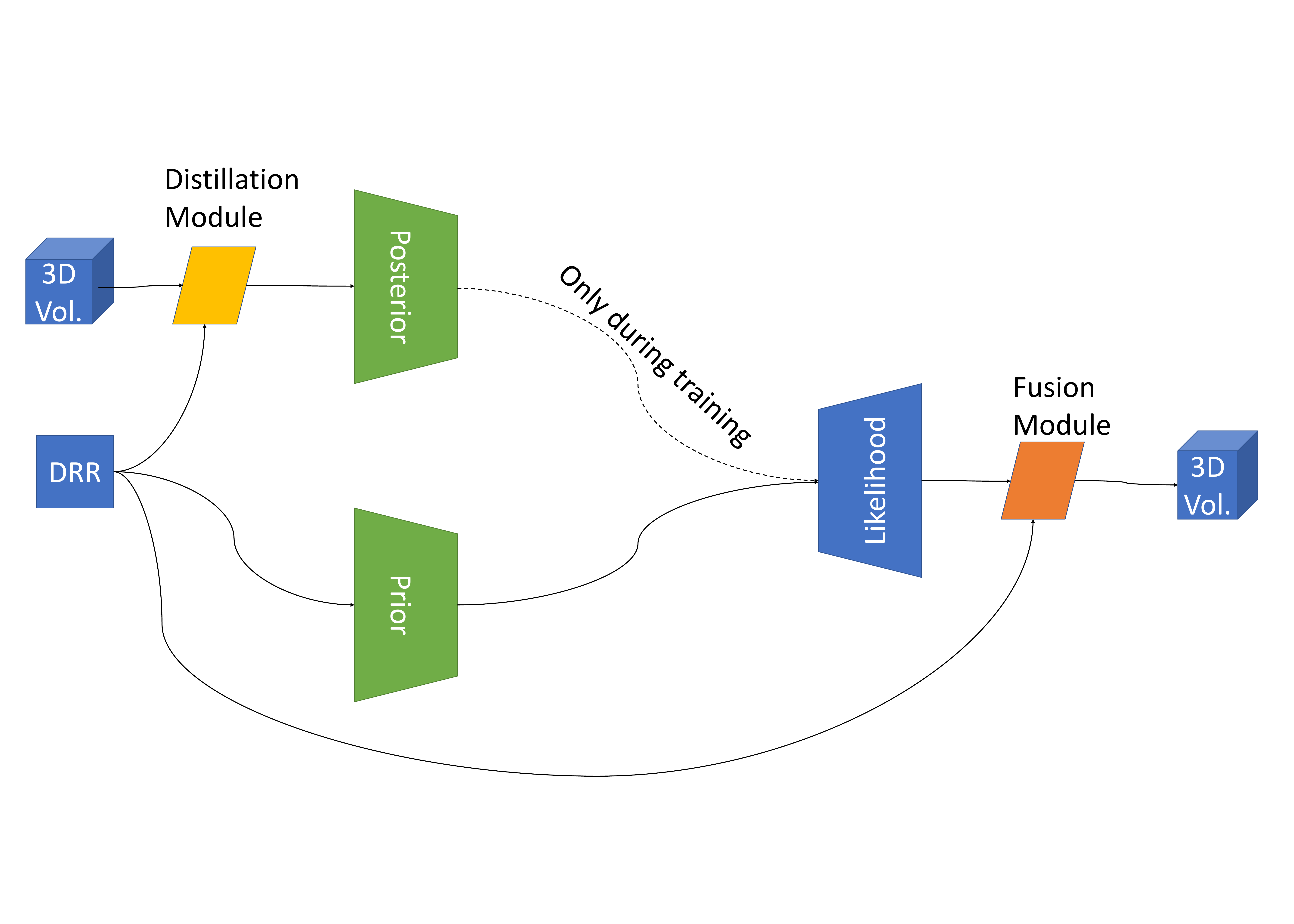}
    }
    \caption{Two approaches for probabilistic 2D-3D un-projection.}
    \label{fig:overview}
\end{figure*}

Adapting a known 2D or 3D architecture to be able to perform a task across dimensions is not a trivial task. Special consideration has to be given in the flow and augmentation of information. As projection is an information destroying process our methods have to be able to deduce the lost information in order to revert the process. This can be achieved through appropriate pathways of information through the network. It is impossible to be entirely certain that the restored information is correct as projection is a many to one operation, thus we believe that a probabilistic approach can offer reasonable confidence intervals. We extend two base architectures to perform this task as they our outlined in Fig.~\ref{fig:overview}. 

\vspace{-10mm}

\vspace{-10mm}

\subsection{2D to 3D MC-Dropout-U-Net}
\label{sect:MC}
 Our first proposed method extends the work of \cite{budd2019confident}. Inputs will be first transformed into three dimensional objects using the structural reconstruction module and then passed through a 3D U-Net~\cite{cciccek20163d}. The U-Net is equipped with dropout layers on the decoding path, which are kept active during inference to mimic stochastic behavior. Fig.~\ref{fig:architecture} shows an overview over the proposed architecture. 


\noindent\textbf{Structural Reconstruction Module:} 2D images can be considered as a 3D image with a ``depth'' of one. A series of five 3D transposed convolutional layers, with stride greater than $1$ in the $z$-axis, is used to match the spatial dimensions of the volumetric 3D target. As opposed to bilinear up-sampling we propose to use transposed convolutions due to their theoretically better ability to  learn more complex and non-linear image resizing functions~\cite{dumoulin2016guide}. The network at this stage contains a conceptual representation of the 3D properties of the object. As the 3D properties of the volume are yet to be fine-tuned by the subsequent 3D U-Net, the output of this layer does not hold human-understandable information of the 3D structure. 

\noindent\textbf{3D Segmentation:}
With the input data in correct spatial dimensions, segmentation can  be performed using a 3D U-Net~\cite{cciccek20163d}. Similarly to its well known 2D counterpart, a 3D U-Net follows an encoding-decoding path with skip connections at each resolution. The network consists of four resolution layers; each consisting of two $3\times3\times3$ kernels with strides of $1\times1\times1$, followed by a $2\times2\times2$ max pooling with strides of $2\times2\times2$. Skip connections are used across the encoding and the decoding path, connecting same resolution levels, in order to propagate encoded features at each resolution. A dropout layer is added at the end of the decoder with a dropout probability of $0.6$. These layers are kept active during inference as per the MC-Dropout methodology~\cite{gal2015dropout}. The network is then trained on 2D images with the respective 3D targets for segmentation and a binary cross-entropy loss.


\subsection{2D to 3D PhiSeg}
\label{sect:phiseg}
In \cite{Baumgartner2019} Baumgartner et al. introduce PhiSeg; a probabilistic segmentation algorithm able to capture uncertainty in medical image segmentations. 

Phiseg is comprised of three modules; the prior, posterior and likelihood networks. The algorithm is modeled after a Conditional Variational Auto-Encoder where the posterior and prior networks operate as the encoders producing a series of latent variables $z$ in different resolution levels. The likelihood network operates as the decoder utilizing the different resolution latent variables sampled from a normal distribution to produce segmentations. It is worth noting that the posterior network takes as input the ground truth segmentation and hence its only used during training. An auxiliary KL divergence loss between the distributions of the prior and the posterior is employed to steer the prior network to produce ``good'' latent variables. 

We extend the previous method in three major ways aimed at controlling and augmenting the information contained in the DRR image. 

\textbf{1. Distillation Module:} We propose a distillation module that performs the inverse operation of the Structural Reconstruction Module and we add it as a pre-processing step of the posterior network. The ground truth image is passed through a series of convolutional layers to distill its 3D information to a 2D representation. The resulting feature maps are  concatenated with the input DRR image and passed through the posterior network. Contrary to the aforementioned 2D-3D U-Net PhiSeg is modeled after a VAE, hence the encoded latent distribution is highly susceptible to noisy inputs. In order to avoid the encoding of noise that would change the characteristics of our distribution we chose to work on 2 dimensions during the encoding phase rather than in 3. We would like to note that a fully 3D PhiSeg with a Structural Reconstruction Module as in the 2D-3D U-Net was evaluated but its training was unstable.

\textbf{2. 3D Likelihood network:} We extend the likelihood network to perform 3D reconstruction. The latent variables that the prior/posterior networks produce are transformed into 3D vectors and used as inputs for the likelihood network. We extend the latent vectors using vector operations rather than learning an augmentation to decrease the computational load of the the network. The series of latent variables are then passed through 3D decoder network, sharing the same architecture as the decoder path of the deterministic 3D U-Net. 

\textbf{3. Fusion Module:} Our next extension of PhiSeg comes in form of a fusion module similar to ~\cite{DBLP:journals/cgf/HenzlerRRR18} at the end of the likelihood module. Contrary to~\cite{DBLP:journals/cgf/HenzlerRRR18} our fusion method is fully learned by the model. Features extracted from the input DRR image $x$ are concatenated to the output $s$ of the likelihood network and convolved together to produce $s'$ which serves as the final output of the network. The intuition behind this module lies with the assumption that PhiSeg will be able to reconstruct the overall structure but may lack details, thus the input DRR image is passed through a convolutional layer to extract relevant features which are then used in conjunction with the proposed segmentation $s$. We also note that the fusion module is not included in the 2D-3D U-Net as the direct skip connections of the model satisfy the flow of information that the fusion module aims at creating.

\textbf{4. Unsupervised Domain Adaptation:}
Finally we propose a new augmentation of PhiSeg aimed at performing Unsupervised Domain Adaptation through self supervision. We chose the task of reconstruction as an auxiliary task in accordance with~\cite{sun2019unsupervised} since it is semantically close to our target segmentation task. To this end we make a new copy of the prior/posterior and likelihood networks that share the weights of the aforementioned modules. We train the resulting model for both segmentation and reconstruction in parallel. Hence the shared encoding paths learn to extract useful information from both domains. In section \ref{exp3} we exhibit results using this technique to segment lungs from NIH X-Rays.

\section{Experiments and Results}
\label{sect:exp}
For our experimentation we focus on two tasks, segmentation and volumetry. 
Two datasets have been used: 60 abdominal CT images of healthy human patients (\emph{Exp1}), and 57 CT porcine livers~\cite{5e59c0fbdf344ad09891ab20bb0a9d6d} (\emph{Exp2}).
Both datasets are resampled to isotropic spacing of $1\textrm{mm}\times1\textrm{mm}\times1\textrm{mm}$. DRRs $\mathbf{p}$ were then generated by projecting the 3D volume on the DRR plane $\mathbf{p}=\mathbf{Mf}$ according to:
$$
p_{i,j}=\sum_{i,j,k}m(i,j,k)f(i,j,k)
$$
where $\mathbf{f}$ is the voxel density vector and $\mathbf{M}$ is the projection matrix calculated using the Siddon-Jacob's Raytracing algorithm~\cite{itk_implementation}.
The synthetic X-ray images of the thorax and porcine abdomen are taken at a fixed distance of 2m and 1m respectively from the CT volume's isocenter, pixel spacing is $0.51mm$. Images contain $512\times512$ pixels, which in this particular configuration, aligns the DRR image and CT volume spatially in pixel space. Both images were then downsampled to $64\times64$ for network training, with the CT volume target centre cropped to preserve spatial alignment with respect to the DRR input.
A third dataset (\emph{Exp3}), obtained from the NIH Clinical Center, was used for a qualitative ablation study. 100 random chest X-ray images from the ChestX-ray8~\cite{nih_test} dataset were selected. No ground truth is present for this experiment.

\textbf{\emph{Exp1}: Compact Structures}
The first experiment assess the network's ability to segment large connected regions. The thoracic CT  dataset was used, with data split; 50 volumes for training and 10 for testing. Annotations were manually made to create ground truth masks for the lung structures. As the lungs appear much darker than other body structures, direct regression to the CT volume is a comparable ground truth target to the manual segmentation masks.

All networks are trained using the Adam-optimizer with an initial learning rate of $1\times10^{-4}$ and a batch size of four. The resulting segmentations were post processed by thresholding based on their pixel intensity values followed by median filtering with a kernel size of $3\times3$ to eliminate sparse noise. 

Table~\ref{tab:DICE-lung-seg} shows the average Dice similarity coefficient (DSC) for the predicted volume compared to the target volume. Dice accuracy for both approaches give equivalent performance. Table~\ref{tab:DICE-lung-seg} also exhibits the ratio between the predicted volume of the lungs and the ground truth. This is is achieved by counting the pixels that lay inside the segmented volume. In terms of quantitative evaluation our deterministic model achieves high Dice score. Meanwhile our dropout and dropblock probabilistic approaches provide us with an on par or better performance to the deterministic method. The variance exhibited on a per sample basis is $0.02$ on dice score and $0.03$ on the ratio of lung capacity. The probabilistic method provides us with more informative lower and upper bound. As the process of projection inherently destroys information, it is our belief that providing an informed upper and lower bounds of our metrics is a more suitable approach.

\begin{figure}[htb]
 \centering
  
  \subfloat[Lung Segmentation with 2D-3D Unet; Left-Right: Input DRR; 3D Ground Truth; 3D Prediction 2D-3D-Unet; Prediction 2D-3D PhiSeg]{
    \includegraphics[height=2.5cm]{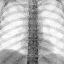} 
    {\includegraphics[height=4.0cm]{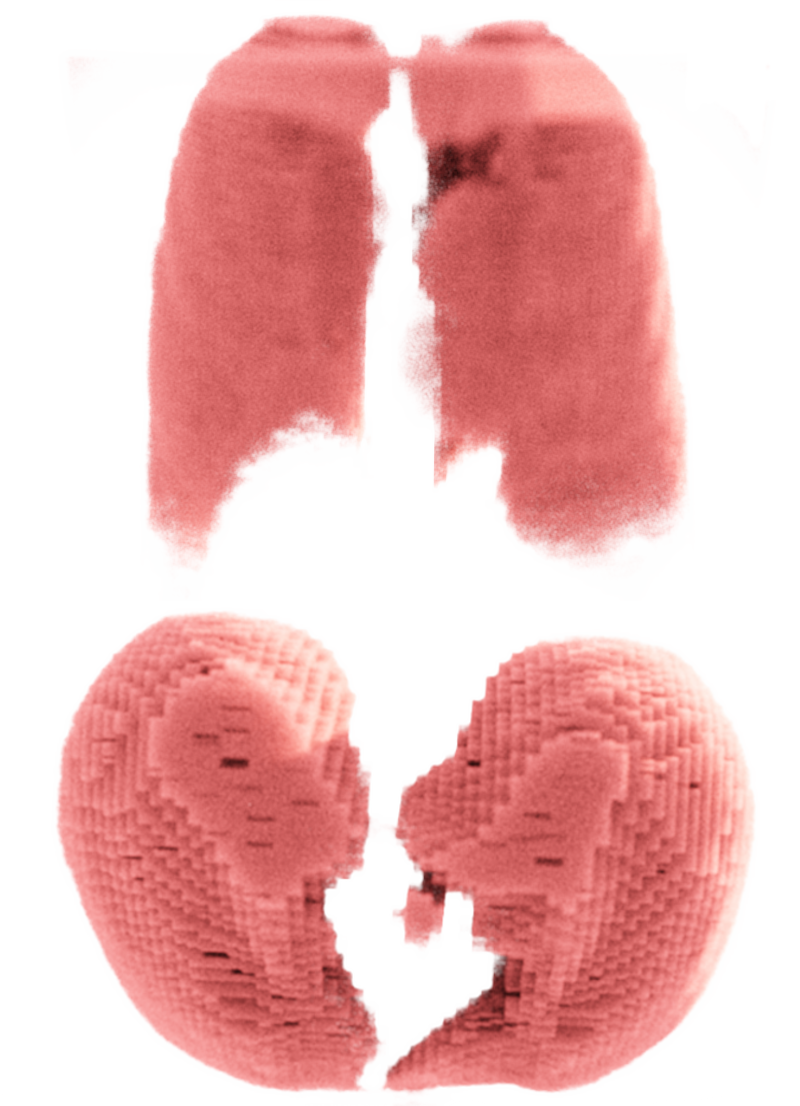} }
    {\includegraphics[height=4.0cm]{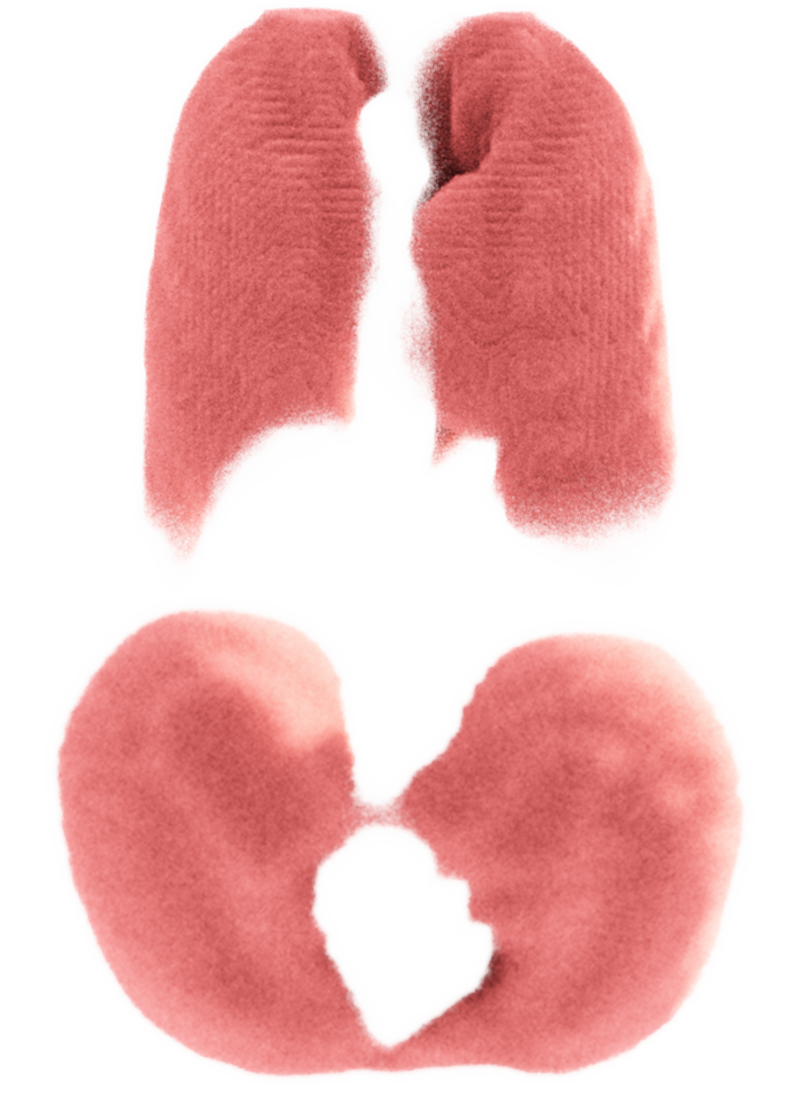}} 
    {\includegraphics[height=4.0cm]{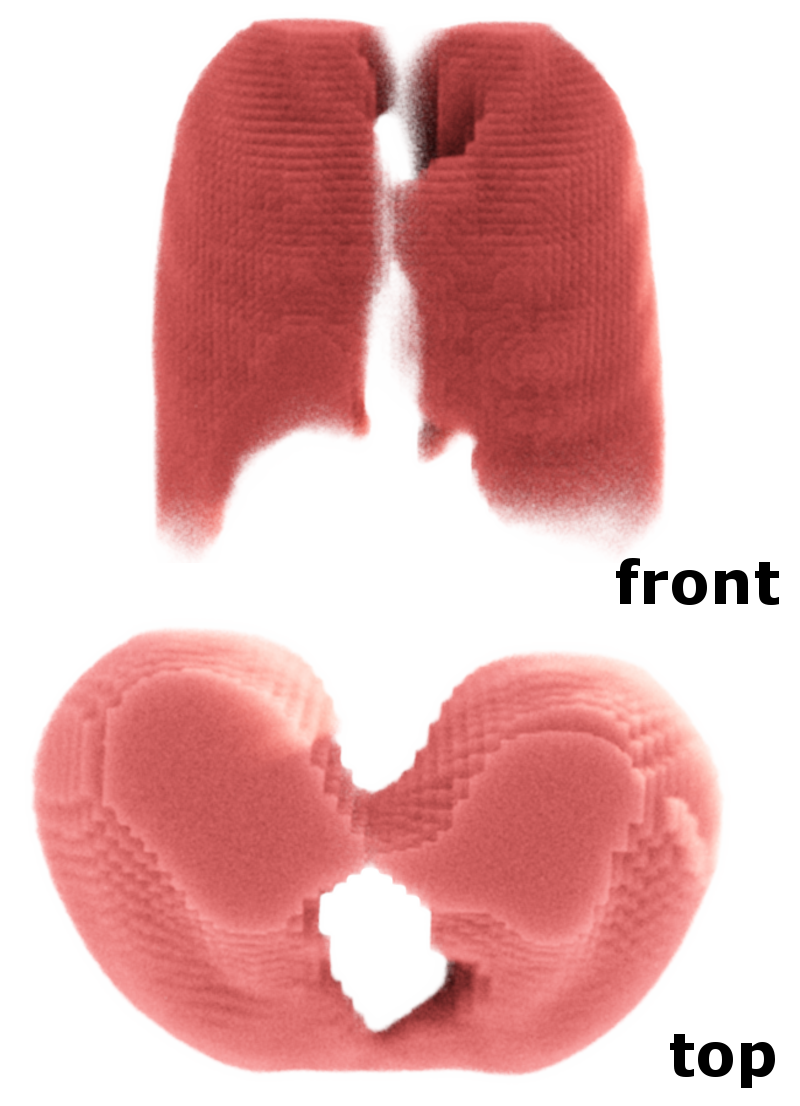}}  
    } \hspace{0mm}
 \subfloat[Porcine Rib Cage Segmentation with 2D-3D Unet; Left-Right: Input DRR; 3D Ground Truth; 3D Prediction 2D-3D-Unet; Prediction 2D-3D PhiSeg]{
    \includegraphics[height=2.5cm]{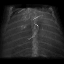} 
    {\includegraphics[height=2.7cm]{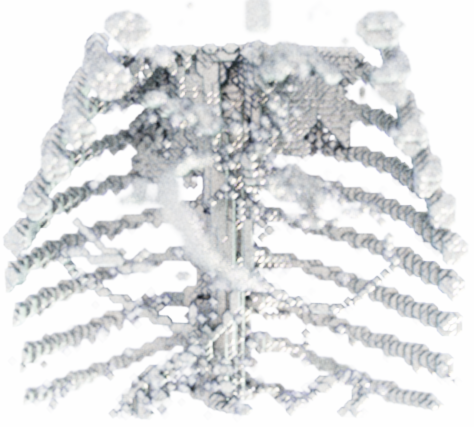} }
    {\includegraphics[height=2.7cm]{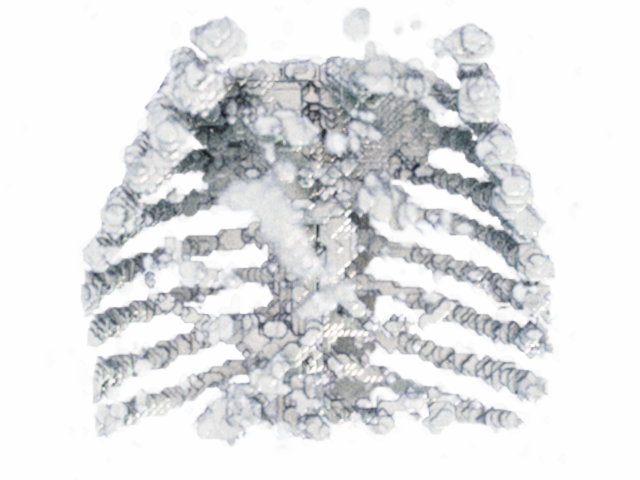}} 
    {\includegraphics[height=2.7cm]{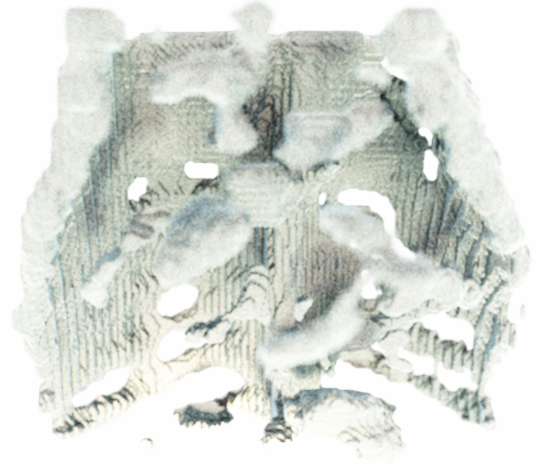}}  
    }

  \caption{Reconstructed Samples for Experiments 1,2. We use~\cite{kroes2012exposure} to enhance depth perception in the 3D figures. }
  \label{fig:exp1}
  \vspace{-6mm}
\end{figure}

Furthermore a version of Phi-Seg without our proposed fusion module was evaluated and noticed a significant increase in the variance of our measurements as well as degraded performance. This observation is in accordance with our hypothesis that the fusion model inserts high level details to our proposed segmentation. 
Qualitative examples are shown in Figure~\ref{fig:exp1}(a).


\begin{table}[]
\centering
\vspace{-6mm}
\begin{tabular}{|l|l|l|l|}
\hline
    \textbf{Method} & \textbf{Exp.1 Volume Ratio} & \textbf{Exp.1 Dice}& \textbf{Exp.2 Dice} \\\hline 
 \textbf{Det. U-Net} & $0.96$ & $0.86$ & $0.41$    \\ \hline
 \textbf{2D-3D PhiSeg} & $0.92\pm0.12$& $\mathbf{0.91\pm0.01}$ & $0.46 \pm 0.05$  \\ \hline
 \textbf{2D-3D PhiSeg w/o fusion} & $1.31\pm0.22$& $0.81\pm0.05$  & $0.45 \pm 0.07$ \\ \hline
 \textbf{2D-3D U-Net Dropout} &$0.91 \pm 0.01$& $0.90\pm0.01$  & $\mathbf{0.48 \pm 0.03}$\\ \hline
 \textbf{2D-3D U-Net Dropblock} &$\mathbf{0.97 \pm0.012}$& $0.83 \pm 0.007$  &  $0.36\pm0.12$\\ \hline
\end{tabular}

 \caption{Average Dice score and Volume Ratio ($\frac{{Predicted}}{{True}}$) of lung and porcine segmentations compared to manually generated 3D ground truth. Exp.2 shows the performance of our methods when the target task is to reconstruct fine 3D structures. }
 \label{tab:DICE-lung-seg}
   \vspace{-10mm}
\end{table}



\textbf{\emph{Exp2}: Fine Structures}
In order to test for limitations and to evaluate the network's ability to segment fine structures we aim to segment the ribcage with the publicly available porcine CT dataset~\cite{5e59c0fbdf344ad09891ab20bb0a9d6d}. Porcine ribs are smaller and finer than human ribs and they project largely on different anatomy (stomach and liver). This data has higher resolution and anatomical focus than the dataset in \emph{Exp 1}, which serves as additional robustness test. The dataset consists of 58 volumes and has been split into 48 volumes for training and 10 for testing. 
Automated thresholding via pixel intensity was used to provide a manual ground truth from the 3D volumes. Known Hounsfield units (HU) for bones in CT have been used to define this threshold ($+1800$ to $+1900$ HU).
The network has been trained with a binary cross-entropy loss, using the Adam optimizer with an initial learning rate of $1\times10^{-4}$ and a batch size of four. Similarly to \emph{Exp1}, the input to the network is a two dimensional DRR image while the segmentation target is the 3D segmentation mask. 

The resulting segmentations achieved an average Dice score of $0.41$ in the deterministic case. Meanwhile our probabilistic approaches where on par or better than the deterministic, achieving a Dice of $0.48\pm0.03$, while providing us with a more informed inference. We note that the difference between our PhiSeg model with and without the fusion module is smaller but still present. We believe this is due to the much harder task of segmenting fine structures across dimensions. Furthermore the Dice score is highly influenced by small outliers caused by noise and a blurry reconstruction of the spine as well as a slight misregistration between the predicted and ground truth volumes. Qualitative results are provided in Fig.~\ref{fig:exp1}(b). Note that small and fine structures as the tips of the ribs are reasonably well formed and shown in the predicted volume.



\textbf{\emph{Exp3}: Domain Adaptability} \label{exp3} 
In order to evaluate the nature of knowledge acquired by the network, and to test potentially limited domain invariance, the network that has been used and trained for \emph{Exp1} was evaluated with  chest X-ray images from the NIH chest X-Ray dataset~\cite{nih_test}. In addition we  evaluate our UDA method on the Montgomery Chest X-Ray dataset that is comprised of 2D chest X-Rays with corresponding 2D segmentations. 

\begin{figure}[htb]
 \centering
  
  \subfloat[UDA on NIH Chest X-Ray Dataset; Left-Right: Input X-Ray; Mean of predicted volume across z-axis; 3D reconstruction of volume.]{
    \includegraphics[height=1.8cm]{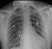} 
    {\includegraphics[height=1.8cm]{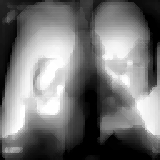} }
    {\includegraphics[height=1.8cm, width=2cm
    ]{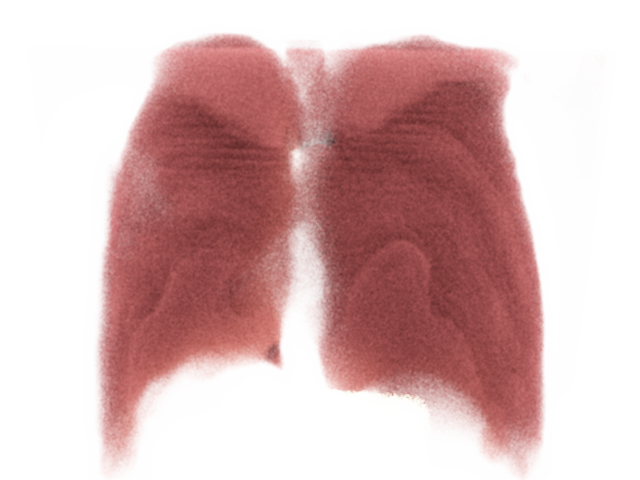}} 
    } \hspace{0mm}
 \subfloat[UDA on the Montgomery X-Ray Thorax Dataset; Left-Right: Input X-Ray; 2D Ground Truth; 2D-3D PhiSeg volume projected onto 2D]{
 
    \includegraphics[height=1.8cm]{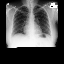} 
    {\includegraphics[height=1.8cm]{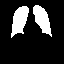} }
    {\includegraphics[height=1.8cm]{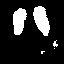}}  
    }
\vspace{-3mm}
  \caption{Examples from Experiment 3.
  }
  \vspace{-5mm}
  \label{nih_fig}
\end{figure}

As it can be seen from Figure~\ref{nih_fig}(a), where the lungs are semi-occluded by a imaging artifact, our network produce the underlying 3D segmentation. This observation signifies that the network learns to reconstruct the anatomy rather than learning a mean lung segmentation. Without corresponding CT volumes, it is not possible to quantitatively evaluate the performance of the network. However, qualitative assessment of 91 subjects shows robust performance of our approach. In the Montgomery dataset the resulting 3D segmentation perimeters are unknown. Thus, we learn a projection to 2D and then compare to the ground truth, resulting in a dice score of \textbf{0.77} when we optimize towards the main DRR-CT task and \textbf{0.86} when we optimize towards the UDA. It is important to note that information is lost during the projection from 3D to 2D during the evaluation period, which explains the decreased performance. In Figure~\ref{nih_fig}(b) we show a selected example from the UDA algorithm.
\vspace{-7mm}

\noindent\textbf{Discussion:} As shown in \emph{Exp1}, our proposed method achieves good Dice scores for 3D lung segmentation while providing informative uncertainty as lower and upper bounds of the volume and dice score.
To the best of our knowledge, this is the first probabilistic methods to perform cross-modality 3D segmentation by unprojecting 2D X-ray images with acceptable performance. 
\emph{Exp2} and \emph{Exp3} have been designed to test expected limitations. In \emph{Exp2} we observe that the prediction of fine structures can work, but with varying performance for either of the methods. \emph{Exp3} shows that our method has promising domain adaptation properties.  However, fine-tuning and calibration will be needed for applications. 


\section{Conclusion}

In this paper we have introduced simple methods to perform probabilistic 3D segmentation from a projective 2D X-ray image. Our networks are data efficient as they have been trained with approximately 60 training DRR-CT pairs and time efficient as they converge within $\sim2$ hours.
In future work we will explore the capabilities of our approach for the reconstruction of vessel trees, e.g. coronary arteries from C-Arm Fluoroscopy. We expect that such reconstructions can be well suited to accurately initialize the registration of pre-operative scans.

\bibliographystyle{splncs03}
\bibliography{references}

\end{document}